\newcommand{\rev}[1]{\textcolor{black}{#1}}

\documentclass[10pt]{article}
\usepackage[utf8]{inputenc}
\usepackage{newunicodechar,graphicx}
\DeclareRobustCommand{\okina}{%
  \raisebox{\dimexpr\fontcharht\font`A-\height}{%
    \scalebox{0.8}{`}%
  }%
}
\newunicodechar{ʻ}{\okina}
\usepackage[letterpaper]{geometry}
\usepackage{hicss}
\usepackage{times}
\usepackage[none]{hyphenat}
\usepackage{booktabs}
\usepackage{soul}
\usepackage{url}
\usepackage[bookmarks=true,hidelinks]{hyperref}

\usepackage{listings}

\usepackage{latexsym}

\usepackage{graphicx} 
\usepackage{longtable} 
\usepackage{xcolor}
\usepackage{tcolorbox}
\usepackage{placeins}
\usepackage{float}
\usepackage{caption}

\floatstyle{plain}
\newfloat{myquote}{H}{lop}[section]
\floatname{myquote}{Quote}
\usepackage{indentfirst}
\usepackage{anyfontsize}
\graphicspath{{images/}}

\lstdefinestyle{mystyle}{
    backgroundcolor=\color{gray!10},   
    commentstyle=\color{green!40!black},
    keywordstyle=\color{blue},
    numberstyle=\tiny\color{gray},
    stringstyle=\color{orange},
    basicstyle=\ttfamily\footnotesize,
    breakatwhitespace=false,
    breaklines=true,
    captionpos=b,
    keepspaces=true,
    numbersep=5pt,
    showspaces=false,
    showstringspaces=false,
    showtabs=false,
    tabsize=2
}

\usepackage[
    style=apa,
  ]{biblatex}
\usepackage{background}
\backgroundsetup{
  position=current page.east,
  angle=-90,
  nodeanchor=east,
  vshift=-4mm,
  opacity=0.5,
  scale=3,
  contents=PREPRINT
}

\newcommand{\RQA}{\textbf{RQ1}: How have mobile app accessibility questions on Stack Overflow grown over the years?}

\newcommand{\RQB}{\textbf{RQ2}: What are the characteristics of mobile app accessibility questions on Stack Overflow?}

\newcommand{\RQC}{\textbf{RQ3}: What are the challenges associated with mobile app accessibility development?}

\DeclareCaptionType{Quote}

\title{Exploring Accessibility Trends and Challenges in Mobile App Development: \\A Study of Stack Overflow Questions}

\author{Amila Indika \\
 University of Hawaiʻi at Mānoa \\
 {\underline{\href{mailto:amilaind@hawaii.edu}{amilaind@hawaii.edu}} }\\ \\
 Justin Lisoway \\
 University of Hawaiʻi at Mānoa \\
 {\underline{\href{mailto:jlisoway@hawaii.edu}{jlisoway@hawaii.edu}} } \\ \And
 Christopher Lee \\
 University of Hawaiʻi at Mānoa \\
 {\underline{\href{mailto:clee48@hawaii.edu}{clee48@hawaii.edu}} } \\ \\
 Anthony Peruma\\
 University of Hawaiʻi at Mānoa \\
 {\underline{\href{mailto:peruma@hawaii.edu}{peruma@hawaii.edu}} } \\ \And
 Haochen Wang \\
 University of Hawaiʻi at Mānoa \\
 {\underline{\href{mailto:hwang400@hawaii.edu}{hwang400@hawaii.edu}} } \\ \\
 Rick Kazman\\
 University of Hawaiʻi at Mānoa \\
 {\underline{\href{mailto:kazman@hawaii.edu}{kazman@hawaii.edu}} } \\ }

\date{}

\begin{document}
\maketitle
\begin{abstract}
The proliferation of mobile applications (apps) has made it crucial to ensure their accessibility for users with disabilities. However, there is a lack of research on the real-world challenges developers face in implementing mobile accessibility features. This study presents a large-scale empirical analysis of accessibility discussions on Stack Overflow to identify the trends and challenges Android and iOS developers face. We examine the growth patterns, characteristics, and common topics mobile developers discuss. Our results show several challenges, including integrating assistive technologies like screen readers, ensuring accessible UI design, supporting text-to-speech across languages, handling complex gestures, and conducting accessibility testing. We envision our findings driving improvements in developer practices, research directions, tool support, and educational resources.
\end{abstract}

\subsubsection*{Keywords:}

mobile apps, Android, iOS, accessibility, stack overflow

\section{Introduction}
With billions of people currently using smartphones for tasks beyond communication---such as finance, health, and entertainment (\cite{smartphone_growth})---it’s essential for mobile application (app) developers to create apps that cater to a wide range of users, including the 1.3 billion people globally with disabilities (\cite{WHO_disability_report}).

Prior research has explored mobile app accessibility in multiple ways, such as analyzing app codebases for accessibility issues (\cite{eler2018automated, Silva2020}), constructing accessibility taxonomies and guidelines (\cite{ballantyne2018study, El-Glaly2018}), construction of tools (\cite{Silva2018, Siebra2018}), and examining user reviews for accessibility feedback (\cite{Reyes2022}). However, there is little research on the real-world challenges in implementing accessibility features. To this end, we conducted a large-scale empirical study of mobile accessibility discussions on Stack Overflow, mining 15 years of data. Stack Overflow is a leading programming Question-and-Answer website with millions of questions, answers, and users (\cite{stackoverflow_survey}).  

\subsection{Goals \& Research Questions (RQs)}
Employing both quantitative and qualitative analyses of question metrics and content, this study aims to understand how extensively mobile app developers seek assistance in incorporating accessibility features into their apps and the key trends, topics, and challenges therein. We address these objectives through the following research questions (RQs):

\vspace{2mm}
\noindent\textbf{\RQA}
This RQ aims to understand how often developers seek help from the community to make their mobile apps accessible. We achieve this by measuring the yearly growth of mobile accessibility questions on Stack Overflow.

\vspace{2mm}
\noindent\textbf{\RQB} 
This RQ involves analyzing the metadata associated with mobile accessibility questions. Specifically, we explore popular tags and gather statistical measures for attributes such as the score, views, answers, comments, and response times. By analyzing these metrics, we can gather insights into the level of engagement and the community’s interest in mobile accessibility topics. 

\vspace{2mm}
\noindent\textbf{\RQC} 
Through this RQ, we aim to identify the range of mobile accessibility challenges app developers discuss. We achieve this by combining natural language processing techniques and manual reviews.

The findings of this study can benefit various stakeholders. App developers can gain insights into common accessibility challenges and better prepare when building their apps. Researchers can use this data to advance the knowledge and understanding of mobile app accessibility. And accessibility advocates and organizations can tailor their support and resources to address the challenges identified.

\section{Related Work}
\label{Section:related}

This section reports on research studies that have examined mobile app accessibility. 

\subsection{Stack Overflow Mining for Mobile Accessibility}
\label{subsection: SO_mining}

\rev{
\textcite{vendome2019can} mined 810 Stack Overflow discussion threads and 13,817 GitHub Android repositories to assess developers' use of assistive technologies, such as Accessibility-APIs. They found that only 2.08\% of apps imported accessibility APIs and were often used for non-accessibility purposes, like notification retrieval and touch interaction automation. Unlike their study, which focused solely on Android and accessibility APIs, our research examines a broader range of iOS and Android questions, focusing on mobile accessibility in general.
\textcite{ma2022first} analyzed dark mode accessibility issues in Android apps by examining posts and over 6,000 apps. They used Latent Dirichlet Allocation (LDA) to categorize developer challenges, summarizing accessibility issues specific to dark mode. However, our study focuses on accessibility issues beyond dark mode.
\textcite{fontao2018supporting} mined over 1.5 million questions related to Android, iOS, and Windows, categorizing discussions into hot topics and unanswered questions using LDA. Although they provided ten strategies for developer governance, our study differs by focusing solely on mobile app accessibility with a smaller dataset and excluding the Windows platform.
Similarly, \textcite{rosen2016mobile} analyzed over 13 million Stack Overflow questions to explore mobile development discussions, categorizing them into 40 categories using LDA topic modeling. They identified difficult-to-answer questions like connectivity, graphics, and media/streaming questions. However, none of these categories specifically addressed mobile accessibility. Lastly, \textcite{linares2013exploratory} explored Stack Overflow questions across various platforms like Android, iOS, BlackBerry, and Windows using LDA. The authors found that most developers answered questions on a single platform. Also, out of the 20 topics identified, none were related to mobile accessibility.
}

\subsection{Repository Mining for Mobile Accessibility}

\rev{
\textcite{ross2020epidemiology} analyzed 9,999 free Android apps, identifying seven accessibility barriers affecting impaired individuals. Although their research offers valuable insights, their dataset was not designed explicitly for accessibility analysis, and their approach does not compare with other platforms like iOS. In contrast, our study explores mobile accessibility across iOS and Android by analyzing Stack Overflow discussion posts, providing a broader view of developers' challenges.
\textcite{di2022making} found that developers lack effective mechanisms to address mobile accessibility issues despite growing interest. Their study examined the implementation of accessibility standards, highlighting a lack of practical guidelines, a lack of developer awareness of disabilities, and technical challenges in implementing accessibility features. Unlike their focus on Android, our study covers iOS and Android and uses Stack Overflow data mining to explore accessibility issues.
\textcite{acosta2020accessibility} identified a significant gap in mobile developers' awareness of accessibility, revealing non-compliance with Web Content Accessibility Guidelines (WCAG) 2.1 in popular apps through manual and automated reviews. Similarly, \textcite{alshayban2020accessibility} studied over 1,000 Android apps, uncovering 11 common accessibility issues. They also surveyed Android developers to understand their perceptions of mobile accessibility. Their research attributes many of these issues to developer unawareness, additional costs, and insufficient support from management, highlighting the challenges disabled users face in finding accessible apps.
}

\rev{
\subsection{Automated Accessibility Detection and Evaluation Tools}
}
\rev{
Automated tools for detecting and repairing accessibility issues are well-documented in existing literature. For example, \textcite{alotaibi2021automated} developed a tool that refactors and repairs identified accessibility problems. Additionally, tools like Xbot (\textcite{chen2021accessible}) focus on detecting accessibility issues, while dynamic test generation tools like MATE (\textcite{eler2018automated}) uncover these issues during testing.
}

\rev{
Existing research has utilized accessibility evaluation tools to assess mobile applications' compliance with accessibility guidelines (\textcite{mateus2021systematic}). Additionally, current literature shows automated accessibility evaluation tools are developed specifically for mobile apps (\textcite{silva2018survey}).
}

\section{Method}
\label{Section:experiment_design}

Our research involved extracting Stack Overflow posts based on their tags and then analyzing the posts' metadata and content. \rev{As of August 2024, Stack Overflow contains approximately 1.42 million questions tagged with ``Android''  and 0.69 million tagged with ``iOS,'' demonstrating the widespread appeal of this platform among mobile app developers.} Our analysis included \rev{mixed-methods approach comprising} quantitative and qualitative techniques. Quantitative techniques involved using standard statistical methods and applying  Top2Vec (\cite{angelov2020top2vec}) topic modeling. Qualitative techniques involved a manual review of a statistically significant sample of the data by the authors. We will provide a detailed explanation of our analysis approach as we discuss each research question. 
\rev{We have made the study's Python scripts, SQL queries, shell scripts, and dataset available.\footnote{\url{https://zenodo.org/doi/10.5281/zenodo.13753236}}}.

\subsection{Dataset}
\rev{The data for this study was collected from Stack Overflow using the Stack Exchange Data Explorer\footnote{\url{https://data.stackexchange.com/stackoverflow}} by querying the \textit{Posts} table. In addition to the keyword \emph{‘accessibility’}, we conducted a snowballing analysis by reviewing tags associated with accessibility to identify additional keywords such as \emph{‘screen readers,’} \emph{‘talkback,’} and \emph{‘voiceover,’}. This approach helped us gather more accessibility-related posts. We then used these newly identified keywords to perform additional queries to gather relevant data.}

\rev{The extraction process involved several steps. First, we filtered accessibility-related questions by extracting post IDs using keywords in the title and tags. Next, we retrieved the question's body, metadata, and answers for the extracted question ID. Finally, we merged the questions and answers from each keyword-related query to create a comprehensive SQLite database of posts; we utilized the ID field to ensure no duplicate posts. The range of dates for the extracted posts was from January 2009 to October 2023.}
We extracted \textbf{6,371 posts}, comprising \textbf{3,022 questions} and \textbf{3,349 answers}. Each record in the database corresponds to a specific post and features 23 distinct fields.

\section{Results}
\label{Section:experiment_results}

\subsection*{\RQA}

\noindent\textbf{Approach:} \rev{RQ1 analyzed the temporal evolution of mobile accessibility challenges by examining the growth of mobile accessibility questions on Stack Overflow over the years. Specifically, we examined the yearly trends in questions without answers, questions without accepted answers, questions with accepted answers, and total questions.}

\begin{figure}[!htb]
    \centering
    \centerline{\includegraphics[keepaspectratio, width=0.40\textwidth]{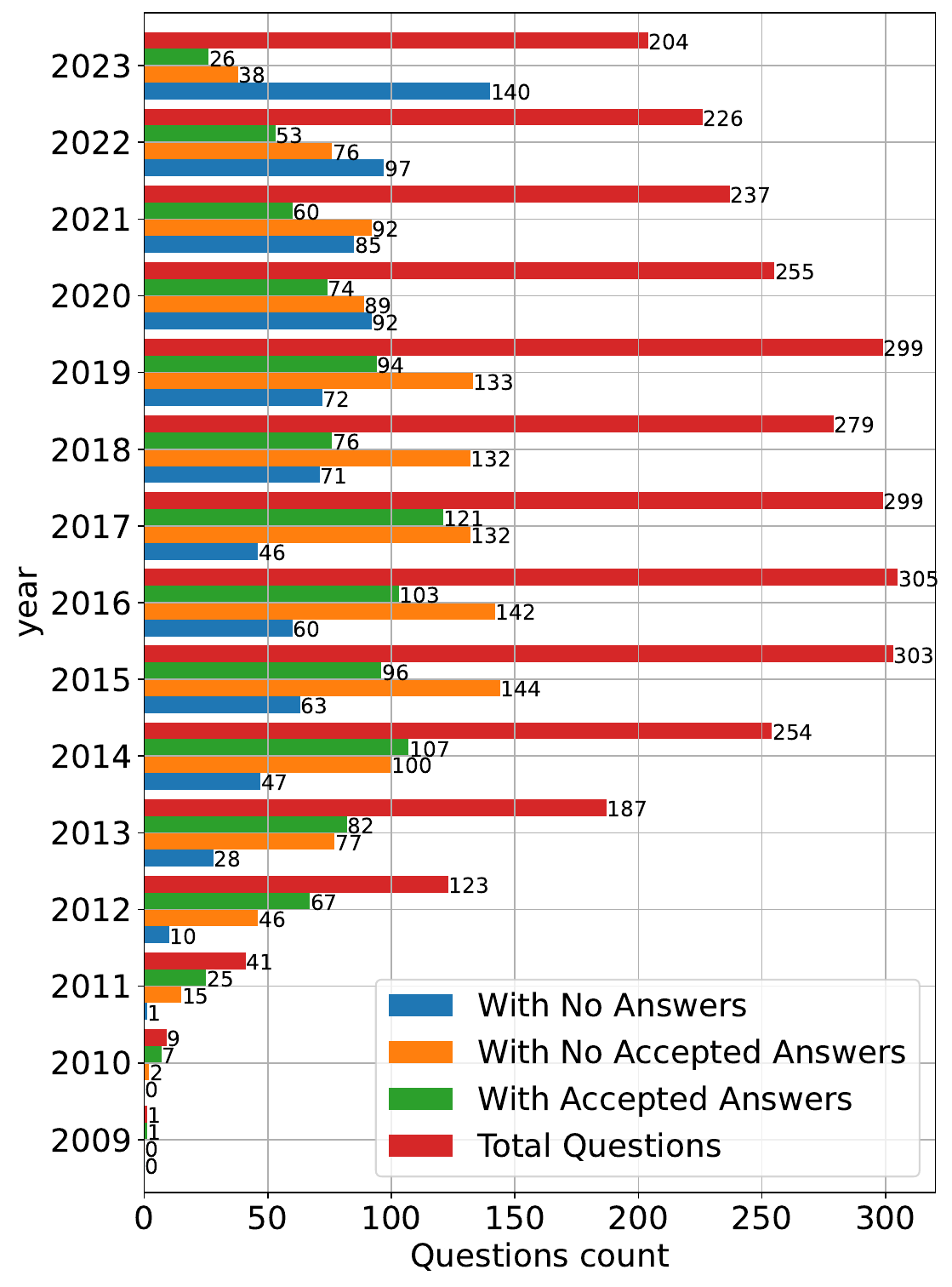}}
    \caption{Mobile Accessibility questions asked by developers over the years}
    \label{fig: question_distribution}
\end{figure}

\rev{Figure \ref{fig: question_distribution} shows this distribution. Out of the 3,022 mobile accessibility questions, 2,210 (73.13\%) received answers, with 992 (32.83\%) having accepted answers and 1,218 (40.30\%) having non-accepted answers. Only 812 (26.87\%) questions remained unanswered. The peak year was 2016, with 305 questions, while the median annual number of questions was 237, with a median of 60 unanswered questions.}

Focusing on Total Questions in Figure \ref{fig: question_distribution}, the early 2010s saw increased discussion. This surge may have been driven by key developments like the introduction of VoiceOver for the iPhone 3GS  (\cite{voiceover_introduction}) and Google’s eyes-free project, which evolved into TalkBack and was integrated into Android 4.0 in 2011\footnote{\url{https://code.google.com/archive/p/eyes-free}}$^{,}$\footnote{\url{https://developer.android.com/about/versions/android-4.0-highlights}}. From 2009 to 2014, our manual analysis identified 228 VoiceOver-related questions ($\approx$7.5\%), such as a question on integrating VoiceOver to support visually impaired users, as noted in a Stack Overflow question (\cite{rq1-example1}). During the same period, there were 106 TalkBack-related questions ($\approx$3.5\%), including a post where a developer sought help with language support compatibility (\cite{rq1-example2}).

In 2011, Android 4.0 also introduced features, such as activating accessibility options through a clockwise rectangle touch gesture, Explore-by-Touch mode, and spoken feedback for screen touches. Developers, however, faced challenges with Explore-by-Touch, as noted in a Stack Overflow question (\cite{android4-challenge}). These challenges are evidenced by over 20 questions ($\approx$1\%) related to explore-by-touch mode from 2009 to 2016. Additionally, Android 4.0 included system-wide font size adjustment, benefiting visually impaired users. Subsequent versions, such as Jelly Bean (4.1, 4.2, and 4.3)\footnote{\url{https://developer.android.com/about/versions/android-4.1\#A11y}}$^{,}$\footnote{\url{https://developer.android.com/about/versions/jelly-bean}} introduced in 2012 and 2013, further improved accessibility features. For instance, version 4.1 introduced the accessibility focus feature for navigation using swipes and double taps, though developers encountered issues with this feature (\cite{android4-challenge2}). Android 4.2 added screen magnification via triple-tapping, zoom-and-pan functionality, and enhanced Braille device feedback (\cite{verge}). The increased questions about gestures, like double taps, swipes, triple taps, and zooming---212 questions ($\approx$7\%) during this period---illustrate growing interest. 
Moreover, mobile app accessibility may have been influenced by regulations. For example, WCAG 2.0, introduced in 2008 (\cite{WCAG_2_0}), offered recommendations for making web content accessible\rev{, which mobile app developers also rely on (\cite{wcag_1, wcag_2}) demonstrate how WCAG is considered when integrating accessibility standards into their apps.}  Also, in 2015, the EN 301 549 European Standard (\cite{etsi_en_301_549}) aimed to conform ICT products and services in Europe with these accessibility requirements.

\rev{We observed a decline in mobile accessibility inquiries after 2016, which might be attributed to increased developer awareness and training. A 2022 survey by \textcite{2022_state_of_digital_accessibility_report} found that 62.7\% of 1,030 participants had received accessibility training, and 91\% used free accessibility tools. This greater understanding and access to resources may have reduced the need for queries.}
Finally, the low number of mobile accessibility-related posts in 2009 and 2010 is likely due to Stack Overflow’s inception in 2008 (\cite{stackoverflow_introduction}).

\begin{tcolorbox}[top=0.5pt,bottom=0.5pt,left=1pt,right=1pt]
\textbf{Summary for RQ1.}
Developers post questions to Stack Overflow when vendors introduce accessibility features, such as VoiceOver and TalkBack, into the operating system. Recently, the number of mobile app accessibility questions on the site has decreased.
The enactment of accessibility regulations and the introduction of new accessibility features by the major vendors are likely factors influencing the growth of accessibility questions. Additionally, developing mature mobile accessibility practices reduces the need for repeated questions.
\end{tcolorbox}

\subsection*{\RQB}

\noindent{\textbf{Approach:}}
RQ2 employs quantitative analysis to examine the characteristics of mobile accessibility questions and answers. First, we extracted tag occurrences in questions. Since our study focuses on mobile accessibility, we excluded the tags `Accessibility,’ `Android,’ and `iOS’ from this analysis \rev{as they did not yield significant insights, considering that these keywords were already involved in the data extraction process.} Next, we calculated the first answer response time by retrieving the creation dates of questions and their first answers and computing the difference in days. Additionally, we queried the ‘ViewCount,’ ‘AnswerCount,’ ‘Score,’ and ‘CommentCount’ per question and counted the number of questions per user. Finally, we computed a statistical summary for these metrics.

\begin{figure}[htbp!]
  \centering
  \includegraphics[keepaspectratio, width=0.45\textwidth]{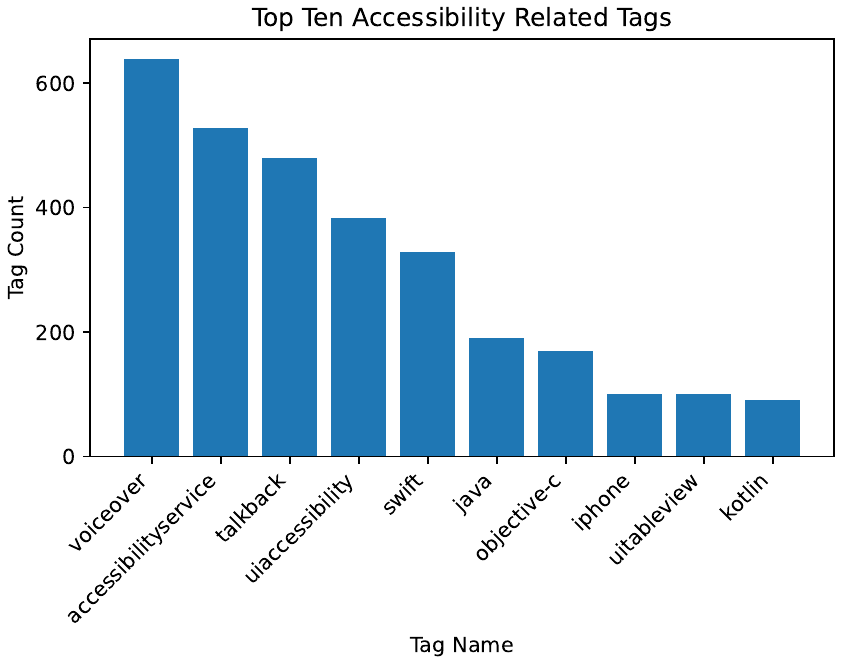}
  \caption{Top ten accessibility-related tags}
  \label{fig: TopTenTags}
\end{figure}

\begin{table*}[!ht]
\centering
\caption{Statistical measures of gathered Stack Overflow posts}
\vspace{-2mm}
\label{tab:statistical_measures}
\begin{tabular}{|l|r|r|r|r|r|r|}
\hline
\multicolumn{1}{|c}{\textbf{Name}} &
  \multicolumn{1}{|c}{\textbf{Average}} &
  \multicolumn{1}{|c}{\textbf{Median}} &
  \multicolumn{1}{|c}{\textbf{StdDev}} &
  \multicolumn{1}{|c}{\textbf{Min}} &
  \multicolumn{1}{|c}{\textbf{Max}} &
  \multicolumn{1}{|c|}{\textbf{Total}} \\
\hline
View Count per Post                      & 1995.31 & 714.00 & 7595.86 & 7.00  & 340916.00 & 6029821.00 \\
First Response Time in days per Question & 99.73   & 1.28   & 320.17  & 0.00  & 3737.42   & 220403.22  \\
Answer Count per Question                & 1.11    & 1.00   & 1.10    & 0.00  & 13.00     & 3349.00    \\
Score per Question                       & 2.61    & 1.00   & 6.66    & -5.00 & 245.00    & 7895.00    \\
Comment Count per Question               & 1.34    & 0.00   & 2.11    & 0.00  & 17.00     & 4054.00    \\
Questions per User                       & 1.27    & 1.00   & 0.97    & 1.00  & 26.00     & 3022.00   \\
\hline
\end{tabular}
\end{table*}

Figure \ref{fig: TopTenTags} shows the top ten accessibility-related tags. We observe that there is interest in screen readers as evidenced by tags such as \emph{voiceover} (i.e., the standard iOS screen reader), \emph{uiaccessibility} (works with \emph{voiceover}), \emph{talkback} (i.e., the standard Android screen reader), and \emph{accessibilityservice} (works with \emph{talkback}). Additionally, \emph{accessibilityservice} and \emph{uiaccessibility} provide accessibility functionality beyond screen readers. Tags related to iOS development languages (\emph{swift} and \emph{objective-c}) and Android development languages (\emph{java} and \emph{kotlin}) were also prominent.

\rev{
Table \ref{tab:statistical_measures} reveals insights into mobile accessibility questions. Using the median, which is robust against outliers, we found that questions are typically answered within two days and receive at least one response, indicating higher community engagement. However, posts often have few comments and limited upvotes. Also, developers tend to ask more questions than they answer, primarily focusing on visual impairments rather than other disabilities.}

\begin{tcolorbox}[top=0.5pt,bottom=0.5pt,left=1pt,right=1pt]
\textbf{Summary for RQ2.}

Mobile accessibility questions on Stack Overflow typically receive responses within two days and at least one answer per query. Authors are more likely to post questions than answers, yet there are more answers than questions overall. Also, these questions generally lack comments and upvotes.
\end{tcolorbox}

\subsection*{\RQC}

\noindent{\textbf{Approach:}}
\rev{In RQ3, we conduct a qualitative and quantitative analysis of mobile accessibility challenges using Top2Vec topic modeling (\cite{angelov2020top2vec}).} The Top2Vec model clusters data to reveal insights into accessibility issues. Similar to \textcite{peruma2022refactor}, we preprocess question bodies using standard natural language processing techniques, including removing redundant spaces, expanding contractions, eliminating HTML tags and URLs, removing non-alphanumeric characters, tokenizing, converting text to lowercase, and filtering out stopwords. We also manually identified and removed custom stopwords. \rev{Then, we fed cleaned unigrams to the Top2Vec model for grouping questions.}
\rev{We manually optimized the Top2Vec model’s hyperparameters, including \emph{$min\_cluster\_size$} and \emph{$min\_samples$} (\cite{hdbscan_params}), and employed the deep-learn option along with parallelized computations for efficiency. We inspected the model’s segregation of unigram tokens to determine the ideal number of topics. With \emph{$min\_cluster\_size=90$} and \emph{$min\_samples=15$}, the Top2Vec model generated meaningful clusters, resulting in seven topics.}

\rev{After clustering, we reviewed a statistically significant random sample from each group (90\% confidence interval, 10\% error margin) for in-depth analysis. Table \ref{tab:question_topic_breakdown} details the number of questions per topic, those manually reviewed per topic, and the most frequent uni-grams per topic. Screen readers and navigation issues are the most common (18.76\%), followed by custom Android accessibility service issues (18.66\%). Thus, significant challenges include integrating screen readers and addressing navigation issues. Other typical problem areas include UI elements and interactions, multilingual support, and dynamic texts. Below, we describe each topic with an example.}

\begin{table*}[!ht]
\caption{Summary of question counts, reviewed questions, and representative words for each Top2Vec topic}
\vspace{-2mm}
\label{tab:question_topic_breakdown}
\scalebox{1}{
\begin{tabular}{|l|r|r|r|l|}
\hline
\textbf{Topic} &
  \multicolumn{1}{l|}{\textbf{\begin{tabular}[c]{@{}l@{}}Question \\ Count\end{tabular}}} &
  \multicolumn{1}{l|}{\textbf{\begin{tabular}[c]{@{}l@{}}Question \\ Percentage\end{tabular}}} &
  \multicolumn{1}{l|}{\textbf{\begin{tabular}[c]{@{}l@{}}Manually \\ Reviewed \end{tabular}}} &
  \textbf{\begin{tabular}[c]{@{}l@{}}Associated  Unigrams\end{tabular}} \\
\hline
Screen Readers and Navigation &
  567 &
  18.76\% &
  61 &
  \begin{tabular}[c]{@{}l@{}}cells, cell, tableview, \\ table, collection\end{tabular} \\ \hline
\begin{tabular}[c]{@{}l@{}}Custom Android Accessibility \\ Services: Configuration \& Deployment\end{tabular}  & 
  564 &
  18.66\% &
  61 &
  \begin{tabular}[c]{@{}l@{}}manifest, service, package, \\ permission, services\end{tabular} \\ \hline
\begin{tabular}[c]{@{}l@{}}Accessibility of UI Elements and UI \\ Interaction\end{tabular} &
  456 &
  15.09\% &
  59 &
  \begin{tabular}[c]{@{}l@{}}edittext, fragment, listview, \\ compose, activity\end{tabular} \\ \hline
\begin{tabular}[c]{@{}l@{}}Multilingual \& Non-Standard Text, \\
and Text-To-Speech (TTS) Support\end{tabular}  & 
  410 &
  13.57\% &
  59 &
  \begin{tabular}[c]{@{}l@{}}impaired, english, language, \\ speech, tts\end{tabular} \\ \hline
Touch gestures and User Interaction &
  397 &
  13.14\% &
  58 &
  \begin{tabular}[c]{@{}l@{}}gestures, perform, touch, \\ finger, fingers\end{tabular} \\ \hline
\begin{tabular}[c]{@{}l@{}}Accessibility Testing, Troubleshooting, \\ and Automation\end{tabular} &
  327 &
  10.82\% &
  57 &
  \begin{tabular}[c]{@{}l@{}}xcode, crash, inspector, \\ uiaccessibility, simulator\end{tabular} \\ \hline
\begin{tabular}[c]{@{}l@{}}Dynamic Text Size in UI Design\end{tabular} & 
  301 &
  9.96\% &
  56 &
  \begin{tabular}[c]{@{}l@{}}font, size, dynamic, \\ large, settings\end{tabular} \\ \hline
\textbf{Total} &
  \textbf{3,022} &
  \textbf{100.00\%} &
  \textbf{411} &
  --------------  \\
\hline
\end{tabular}
}
\end{table*}

\vspace{1mm}
\noindent{\textbf{Screen Readers and Navigation:}} Screen reader support is essential for making digital content accessible to users with visual and motor impairments (\cite{lazar2007frustrates}). Our manual inspection revealed that mobile developers often encounter challenges integrating screen readers with data visualizations like tables and collections. In particular, iOS developers face difficulties with VoiceOver integration for TableViews, CollectionViews, and ScrollViews, while Android developers have issues with Talkback integration for RecyclerViews. An example is shown in Quote \ref{quote: quote_1}.

\begin{myquote}[!htb]
    \centering
    \begin{tcolorbox}[left=1pt,right=1pt,top=1pt,bottom=1pt,colframe=blue!50!black, colback=blue!10, width=0.47\textwidth]
    \small\textbf{UITableView within a UITableViewCell is not selectable by VoiceOver}
    
    \vspace{3mm}
     \small “... Does anyone know how to make a UITableView inside of a UITableViewCell selectable by VoiceOver?”
    \end{tcolorbox}\vspace{-3mm}
    \caption{An iOS developer requesting help with VoiceOver integration to UITableView (\cite{quote_1})}
    \label{quote: quote_1}
\end{myquote}

Another common challenge is integrating screen readers with nested views or subviews. In addition, guiding impaired users to focus on specific accessibility components with screen readers is complex, requiring logical flow, consistency, contextual information, and efficient navigation. Developers also encounter challenges when implementing screen readers to follow specific gesture-based navigation patterns.

\vspace{1mm}
\noindent{\textbf{Custom Android Accessibility Services: Configuration \& Deployment:}}
Android developers face specific challenges when implementing custom accessibility services, particularly during their early learning phase. Issues often arise with permissions on particular brands/models and older devices and configuring accessibility services in the manifest file. Developers also struggle with enabling accessibility service options after a reboot. In addition, questions about compatibility and deployment across different platforms and debugging Android configurations are typical. An example is shown in Quote \ref{quote: quote_2}.

\begin{myquote}[!htb]
    \centering
    \begin{tcolorbox}[left=1pt,right=1pt,top=1pt,bottom=1pt,colframe=blue!50!black, colback=blue!10, width=0.47\textwidth]
    \small\textbf{AccessibilityService stops receiving events on reboot}
    \vspace{3mm}

    \small`` ...
    If you reboot, you have to disable/re-enable the service from the accessibility services menu. Why does the app not get the events after a reboot?”
    \end{tcolorbox}\vspace{-3mm}
    \caption{An Android developer asking help to re-enable AccessibilityService after reboot (\cite{quote_2})}
    \label{quote: quote_2}
\end{myquote}

\vspace{1mm}
\noindent{\textbf{Accessibility of UI Elements and UI Interaction:}} Developers frequently encounter issues providing accessibility for UI interactions and elements (e.g., text, buttons, image views, radio buttons, and text views), as shown in Quote \ref{quote: quote_3}. Challenges include text fields not updating correctly and redundant readings of hints and content descriptions, leading to confusion for users. Focus management problems and overlapping views like popups/dialogs further complicate navigation. Precise announcements of errors and labels and preventing repetitive auditory feedback are essential for enhancing UI accessibility. Moreover, fixing issues like misreading list items and improper focus on UI fragments is crucial to providing a seamless user experience.

\begin{myquote}[!htb]
    \centering
    \begin{tcolorbox}[left=1pt,right=1pt,top=1pt,bottom=1pt,colframe=blue!50!black, colback=blue!10, width=0.47\textwidth]
    \small\textbf{Android Increase touch target of edit text as per accessibility without affecting design}
    \vspace{3mm}

    \small``...
    If I add 14dp padding to top and bottom then the editText is as per accessibility but it is affecting design and increasing height of editText which I don’t want. Is there any alternate solution to increase touch target with affecting design?”
    \end{tcolorbox}\vspace{-3mm}
    \caption{An Android developer seeking help with setting accessibility for an EditText field (\cite{quote_3})}
    \label{quote: quote_3}
\end{myquote}

\vspace{1mm}
\noindent{\textbf{Multilingual \& Non-Standard Text, and Text-To-Speech (TTS) Support:}} Developers often struggle to support multiple languages like Spanish, Chinese, Tamil, German, and Hebrew, as illustrated in Quote \ref{quote: quote_4}.
Interference between TalkBack and Text-To-Speech (TTS) services is also typical. In addition, reading non-standard text such as acronyms, special characters, and mathematical expressions poses difficulties. Contextual ambiguities, like acronyms and homonyms, can also affect TTS output. For example, “no of” can be read as either “number of” or “no of”. Moreover, visually impaired developers face challenges while developing these accessibility features.

\begin{myquote}[!htb]
    \centering
    \begin{tcolorbox}[left=1pt,right=1pt,top=1pt,bottom=1pt,colframe=blue!50!black, colback=blue!10, width=0.47\textwidth]
    \small\textbf{iOS voiceOver/accessibility foreign words pronuncation}
    \vspace{3mm}\vspace{-3mm}

    \small``... 
    The word ‘Sound’ spoken with German VoiceOver language doesn’t make sense (it should say ‘saund’, but sounds like ‘sund’). Is there a way to give voice-over information about a word’s language?”
    \end{tcolorbox}\vspace{-3mm}
    \caption{A developer having difficulty providing screen reader support for German. (\cite{quote_4})}
    \label{quote: quote_4}
\end{myquote}

\vspace{1mm}
\noindent{\textbf{Touch Gestures and User Interaction:}}
Developers often struggle with handling multiple finger gestures, such as swipes, clicks, long presses, sliding, and zoom-and-panning, with accessibility services, as shown in Quote \ref{quote: quote_5}. Common issues include user interaction using gestures such as swiping through lists, controlling seek bars, and accessing keyboard overlays. Additional challenges involve implementing auto-clicking, magic tap with iOS, and screen reader support with explore-by-touch and double-taps. 

\begin{myquote}[!htb]
    \centering
    \begin{tcolorbox}[left=1pt,right=1pt,top=1pt,bottom=1pt,colframe=blue!50!black, colback=blue!10, width=0.47\textwidth]
    \small\textbf{How to dispatch multi touch gestures using AccessibilityService (disptachGesture)}
    \vspace{3mm}

    \small “... I want to develop an app that dispatches complex gesture for user. 
    ... 
    how can I dispatch them using Accessibility Service (dispatchGesture) function.”
    \end{tcolorbox}\vspace{-3mm}
    \caption{An Android developer seeks assistance for integrating multi-touch gestures (\cite{quote_5})}
    \label{quote: quote_5}
\end{myquote}

\vspace{1mm}
\noindent{\textbf{Accessibility Testing, Troubleshooting, and Automation:}}
\begin{myquote}[!htb]
    \centering
    \begin{tcolorbox}[left=1pt,right=1pt,top=1pt,bottom=1pt,colframe=blue!50!black, colback=blue!10, width=0.47\textwidth]
    \small\textbf{OCMock - how to mock accessibility in iOS}
    \vspace{3mm}
    
    \small``I am using xcode and my mocking frameworks is OCMock. How can i use OCMock to mock that accessibility is turned on so i can run some simple accessibility UI tests? ...”
    \end{tcolorbox}\vspace{-3mm}
    \caption{An iOS developer facing an issue with accessibility UI tests (\cite{quote_6})}
    \label{quote: quote_6}
\end{myquote}

\vspace{1mm}
Developers often encounter challenges with accessibility testing and debugging tools, particularly iOS developers who struggle with tools such as Accessibility Inspector, Appium Inspector, XCTest, and KIF. Quote \ref{quote: quote_6} provides an example. These challenges include integrating accessibility tests using XCTest and KIF, comparing accessibility elements between Xcode’s Accessibility Inspector and Appium Inspector, and the complex nature of UI automation and testing for validating accessibility features. Additionally, there are challenges in cross-platform development using Xamarin and implementing custom actions like modifying configurations, inverting colors, or preventing VoiceOver from reading certain elements.

\vspace{1mm}
\noindent{\textbf{Dynamic Text Size in UI Design:}}
Developers frequently face issues with dynamically adjusting text sizes based on accessibility settings, using tools like iOS’s dynamic-type feature or Cascading Style Sheets (CSS) adjustments. They also struggle with retrieving and applying settings for font size and color adjustments. Platform-specific challenges commonly arise with Flutter, iOS, or Android APIs when implementing dynamic text size features. Maintaining UI integrity and aesthetics across elements like buttons, list views, and data grids is also problematic. Quote \ref{quote: quote_7} illustrates an example of such a scenario.

\begin{myquote}[!htb]
    \centering
    \begin{tcolorbox}[left=1pt,right=1pt,top=1pt,bottom=1pt,colframe=blue!50!black, colback=blue!10, width=0.47\textwidth]
    \small\textbf{Limit supported Dynamic Type font sizes}
    \vspace{3mm}

    \small``I want to support Dynamic Type but only to a certain limit
    ...
    Is there an easy way to accomplish this with standard UITableViewCell styles?”
    \end{tcolorbox}\vspace{-3mm}
    \caption{An iOS developer struggles with setting Dynamic Type font sizes (\cite{quote_7})}
    \label{quote: quote_7}
\end{myquote}

Font-related issues are also a concern. Developers need to handle emphasis and ensure fonts scale dynamically. Integrating dynamic text size adjustments with other accessibility features, such as color and contrast adjustments, transparency, and touch gestures, adds another layer of complexity. Cross-platform frameworks like Xamarin and Flutter also present challenges for handling text size adjustments.

\begin{tcolorbox}[top=0.5pt,bottom=0.5pt,left=1pt,right=1pt]
\textbf{Summary for RQ3.}
Developers face challenges when implementing accessibility features. These include dynamically adjusting text sizes while maintaining UI integrity, conducting accessibility testing and debugging, supporting multiple languages and non-standard text for Text-To-Speech, handling complex touch gestures and user interactions, ensuring accessibility of UI elements and interactions, configuring and deploying custom Android accessibility services, and integrating screen readers with data visualizations and navigation.
\end{tcolorbox}

\section{Discussion}
\label{Section:discussion}
\rev{Our research highlights the need to prioritize accessibility in mobile app design and development, ensuring all users can fully engage with mobile technology and benefit equally. By analyzing 15 years of developer questions, we identified a broad spectrum of accessibility barriers, corroborating the insights of previous studies focused on specific app codebases.}

\rev{Our findings are closely parallel with those of \textcite{fontao2018supporting}, as both highlight Stack Overflow's popularity for discussing mobile app issues and observe accessibility questions decline post-2015. We also found that the median response time for mobile accessibility questions is under two days, consistent with their observation that most questions across various tags were answered within 48 hours. Top tags related to mobile accessibility, such as Swift, Java, and UITableView, appeared among the top 30 tags in Fontao et al.'s study. However, other accessibility tags were absent, likely due to our specific focus on mobile accessibility. Additionally, both studies suggest that official events influence question frequency.}

Our RQ3 findings indicate that most mobile developers prioritize support for the visually impaired, often neglecting other impairments, similar to trends observed by \textcite{vendome2019can}. We found that many Android developers struggle with Accessibility Services, explaining the underutilization of Android Accessibility APIs noted by Vendome et al. 
\rev{
\subsection{Theoretical Underpinnings}
The Technology Acceptance Model (\cite{davis1989perceived}) helps explain mobile developers' engagement with accessibility features by focusing on Perceived Ease of Use (PEOU) and Perceived Usefulness (PU). As observed in RQ3, developers recognize the value of accessibility in enhancing app quality and broadening the user base (high PU). However, they struggle with the technical challenges of implementation (low PEOU), often seeking community support. This hinders the adoption of accessibility best practices, even when developers understand their importance. Addressing low PEOU through better tools, documentation, and community support is essential to increasing the adoption of accessibility features in mobile apps.
}

\rev{
\subsection{Takeaways and Implications}
}
\noindent{\textbf{\rev{Need for improved developer education and training on mobile accessibility.}}}
\rev{
Our findings reveal that developers face significant challenges incorporating accessibility features, resulting in low PEOU. This presents an opportunity for educators to offer specialized mobile app courses or training on accessibility, catering to developers of all skill levels. Organizations should also prioritize accessibility-focused training for their developers, which could ultimately increase PEOU and lead to greater adoption of accessibility features.
}

\noindent{\textbf{\rev{Capturing accessibility requirements early in the software development lifecycle.}}}
\rev{
Developers and business analysts should explicitly define accessibility requirements during the requirements-gathering stage to ensure the app supports necessary disabilities. This understanding allows architects and designers to integrate accessibility into the app’s architecture and UI from the outset. This approach allows project managers to involve accessibility specialists early or provide relevant training, minimizing the need for rework later in the development lifecycle. Additionally, insights from our RQ3 findings can help teams anticipate and address common accessibility challenges proactively.
}

\noindent{\textbf{\rev{Advancing accessibility tools through collaborations between tool vendors and researchers.}}} 
\rev{
Our findings, especially from RQ3, illustrate the need for more advanced and user-friendly accessibility tools. Collaboration between academic researchers and tool vendors could drive the creation of intuitive, innovative, practical, and scalable solutions that integrate seamlessly into developers' workflows, making accessibility implementation more efficient. Additionally, accessibility testing frameworks (e.g., Axe) and libraries that simplify accessible component integration can enhance PEOU. When developers are familiar with these tools, they are more likely to adopt accessibility practices.
}

\section{Threats To Validity}
\label{Section:threats}
Although other sources exist for mobile app developer discussions (e.g., Reddit, \rev{Discord}, GitHub, etc.), our study is limited to Stack Overflow, a popular programming Q\&A website. 
We constructed our query using specific terms and limited it to Android and iOS. Hence, even though we extracted 3,022 questions, there is still room for further analysis and expansion. Our qualitative analysis involves a manual review of a statistically significant sample of data. Even so, there is a chance that the reviewed sample might not include all scenarios present in the dataset. Additionally, as our study is scoped to the topics and challenges in mobile app accessibility, our analysis is only limited to questions and not answers, which can be analyzed in future studies. Finally, our findings provide a point-in-time snapshot of the mobile accessibility challenges developers encounter.

\section{Conclusion \& Future Work}
\label{Section:conclusion}
\rev{
As mobile apps become integral to daily life, developers must ensure they are accessible to users with disabilities. This study analyzes over 15 years of Stack Overflow questions, revealing trends and challenges in mobile app accessibility. The findings show a rise in accessibility questions, often coinciding with the introduction of new mobile accessibility features in Android and iOS. Most accessibility questions receive answers within two days. Key challenges include integrating screen readers, managing custom accessibility services, ensuring accessible UI elements and interactions, supporting multi-language text-to-speech, implementing complex gestures, and conducting accessibility testing.
}

\rev{
The insights from this study highlight the importance of considering accessibility throughout the lifecycle, not just as an afterthought. Educators and advocates can use these findings to connect academic curricula with real-world accessibility challenges. Furthermore, our findings can guide the formulation of new guidelines, best practices, and resources to tackle the critical challenges confronting developers.
}

\rev{
Our future work includes surveying professional mobile app developers to validate this study's findings. We will take two approaches: conducting case studies with organizations that build mobile apps to understand their accessibility practices and launching a large-scale online survey to gather insights on accessibility challenges. We plan to identify survey participants through platforms like LinkedIn.
}

\printbibliography

\end{document}